\documentclass[aps,prd,floatfix,nofootinbib,superscriptaddress,twocolumn]{revtex4-1}
\usepackage{graphicx}  
\usepackage{dcolumn}   
\usepackage{bm}        
\usepackage{amssymb}   
\usepackage{amsmath}
\usepackage{comment}
\usepackage{graphicx}
\usepackage{subfig}
\usepackage{color}
\usepackage{slashed}
\usepackage{hyperref}

\newcommand{\Eq}[1]{Eq.(\ref{#1})}
\newcommand{\Eqs}[2]{Eqs.(\ref{#1}-\ref{#2})}

\begin{document}

\author{Niklas Mueller}
\email{nmueller@bnl.gov}
\affiliation{Physics Department, Brookhaven National Laboratory, Bldg. 510A, Upton, NY 11973, USA}
\author{Raju Venugopalan}
\email{raju@bnl.gov}
\affiliation{Physics Department, Brookhaven National Laboratory, Bldg. 510A, Upton, NY 11973, USA}

\title{Constructing phase space distributions with internal symmetries}
\date{\today}
\begin{abstract}
We discuss an ab initio world-line approach to constructing phase space distributions in systems with internal symmetries. Starting from the Schwinger-Keldysh real time path integral in quantum field theory, we derive the most general extension of the Wigner phase space distribution to include color and spin degrees of freedom in terms of dynamical Grassmann variables. The corresponding Liouville distribution for colored particles, which obey Wong's equation, has only singlet and octet components, while higher moments are fully constrained by the  Grassmann algebra. The extension of phase space dynamics to spin is represented by a generalization of the Pauli-Lubanski vector; its time evolution via the Bargmann-Michel-Telegdi equation also follows from the phase space trajectories of the underlying Grassmann coordinates. Our results for the Liouville phase space distribution in systems with both spin and color are of interest in fields as diverse as chiral fluids, finite temperature field theory and polarized parton distribution functions. We also comment on the role of the chiral anomaly in the phase space dynamics of spinning particles. 
\end{abstract}
\maketitle
\section{Introduction}
Classical phase space distributions involving internal degrees of freedom such as spin and color are useful in describing a wide range of physics across energy scales. The construction of classical phase space distributions with internal symmetries through the use of Grassmann variables was pioneered by Berezin and Marinov \cite{Berezin:1976eg}. They demonstrated straightforwardly how one recovers in this framework the Bargmann-Michel-Telegdi (BMT) equation~\cite{Bargmann:1959gz} describing the precession of spins in background fields. Several contemporaneous studies complemented the work of Berezin and Marinov and extended the Grassmann algebra description of internal symmetries to describing the dynamics of both spin and color degrees of freedom~\cite{Balachandran:1976ya,Balachandran:1977ub,Barducci:1976xq,Barducci:1982yw,Brink:1976uf,Barut:1988ee,Barducci:1976qu,Barducci:1980xk}. The Wong equations~\cite{Wong:1970fu,Kosyakov:2007qc} describing the precession of non-Abelian point particles in colored background fields are also recovered in this approach.

The world-line formalism~\cite{Strassler:1992zr,DHoker:1995uyv,DHoker:1995aat,Mondragon:1995va,Mondragon:1995ab,Hernandez:2008db,JalilianMarian:1999xt,Schubert:2001he,Bastianelli:2006rx, Corradini:2015tik,Bastianelli:2013pta} provides an elegant method, from first principles in quantum field theory, to derive classical phase space distributions in systems with internal symmetries. In this approach, one-loop effective actions are expressed by quantum mechanical point-particle path integrals with internal symmetries incorporated through Grassmann degrees of freedom. In this work, we will develop the Schwinger-Keldysh (SK) generalization of the world-line formalism for applications to nonequilibrium physical processes\footnote{A detailed study of its applications in finite temperature quantum field theory will be discussed separately in \cite{longpaper}.}. The classical color phase space limit is obtained explicitly from the saddle point of the world-line effective action and the phase space dynamics of the extended color and spin symplectic algebra is obtained from Grassmannian classical Poisson brackets as well as a Grassmann valued phase space measure.

For colored degrees of freedom, a canonical coordinate transformation from Grassmann variables to Grassmann bilinears relates our approach to that of Alekseev, Faddeev and Shatashvili (AFS) \cite{Alekseev:1988vx}. The AFS approach described the symplectic orbits of compact Lie groups with functional integrals involving phase space Darboux variables. 
In particular, the AFS construction of symplectic structures for internal symmetries contains a classical action with a topological Berry monopole~\cite{Berry:1984jv}. In contrast, no such topological terms arise in our derivation of the classical limit by a saddle-point approximation to the SK world-line path integral. We also demonstrate that the Grassmann algebra allows one to express the Liouville density for colored particles uniquely in terms of two independent color structures. 

One can similarly construct classical phase space distributions in the world-line approach for both massless and massive fermions recovering the results of Berezin and Marinov \cite{Berezin:1976eg}. Here too, a canonical transformation from Grassmann spin variables to their commuting bilinear phase space counterpart, and the corresponding simple rules for the spin phase space measure, generates a unique and elegant form for the classical phase space distribution function for spin in these variables. This derivation provides a many-body generalization of the relativistic  Pauli-Lubanski spin pseudo-vector \cite{Berezin:1976eg}, whose time evolution reproduces the BMT equation. 

One can also define, in this language of Grassmann bilinears, chiral vector and and axial vector currents as classical phase space averages weighted by the generalized Liouville density. We show that a naive generalization of our approach misses the effects of the chiral anomaly. Recovering the anomaly requires a careful treatment whereby one introduces auxiliary axial vector fields in the Schwinger Keldysh world-line path integral.

Our manuscript is organized as follows: In section \ref{sec:Grassman}, we provide a short derivation of classical phase space distributions with internal symmetries from a saddle point expansion of the world-line path integral expression for the one loop effective action in quantum field theory. A more detailed computation will be presented in \cite{longpaper}. We derive in the Schwinger-Keldysh formalism, via a ``truncated Wigner" (or classical-statistical) approximation, the Wigner distribution for the extended symplectic phase space involving Grassmann variables\footnote{General properties of fermionic Wigner functionals are discussed in e.g. \cite{Gao:2017gfq,Ohnuki:1978jv,Abe:1989mr,Mrowczynski:2012ps,Vasak:1987um,Elze:1989un}.}. Our construction involves canonical phase space variables and we will show that the semi-classical limit of quantum phase space is incompressible \cite{Moyal:1949sk,Cary:1983zz,Littlejohn:1991zz}. 

In section \ref{sec:color}, we discuss the structure of the Grassmann valued Wigner phase space for color degrees of freedom and demonstrate that the group algebra is exactly realized in the classical limit for any finite dimensional representation. 
An intuitive description of the extended phase space for color is obtained by expressing our results in terms of Grassmanian bilinear color charges whose dynamics is given by Wong's equations \cite{Wong:1970fu}. This allows us to replace the  Grassmann formulation of the extended Wigner phase space distribution (and the corresponding Grassmann valued measure) by a classical phase space distribution (and phase space measure), whose elements are color charges.

Section \ref{sec:spin} is devoted to the Grassmann description of relativistic spin and chirality. In analogy to the case of color, the spin Wigner function can be replaced by a classical phase space distribution including a many-body generalization of the Grassmanian bilinear Pauli-Lubanski spin pseudo-vector satisfying the BMT equation.

In section \ref{sec:anomaly}, we briefly describe from first principles how the chiral anomaly manifests itself in this phase space approach. We first show that the expectation value of the axial-vector current, constructed from a semi-classical phase space average, is naively conserved. We show that the proper treatment of its violation by the anomaly is obtained by introducing an auxiliary axial-vector gauge field in the path integral; the four-derivative of the corresponding axial vector current is robust when this axial vector field is subsequently put to zero and generates the well known expression for the axial anomaly. 

Our derivation provides a clean and transparent derivation in quantum field theory of classical phase space distributions in terms of internal degrees of freedom that have a simple physical interpretation but are nevertheless fully constrained by the underlying Grassmann algebra. The framework developed here potentially has wide ranging applications in nonequilibrium processes; in our concluding remarks, we briefly outline our work in progress on some of these problems.

Our work is supplemented by appendices: In appendix \ref{app:phasepsace}, we discuss Liouville's theorem and the (in-)\-com\-pressibility of phase space in the semi-classical limit. We provide details of the classical color phase space construction in appendix \ref{app:color}. 
In appendix \ref{app:Berry}, we outline how the incompressibility of phase space may be affected by the presence of a Berry phase~\cite{Berry:1984jv,Sundaram:1999zz,Xiao:2005qw,Son:2012wh, Stephanov:2012ki,Son:2012zy,Duval:2014ppa,Duval:2005vn} and discuss possible ambiguities in its interpretation.
%
%
%
\section{World-line formulation of real-time Quantum Field Theory}\label{sec:Grassman}
We begin by reviewing the world-line computation of one-loop effective actions \cite{Strassler:1992zr,DHoker:1995uyv,DHoker:1995aat,Mondragon:1995va,Mondragon:1995ab,Hernandez:2008db,JalilianMarian:1999xt,Schubert:2001he,Jeon:2004rk,Bastianelli:2006rx, Corradini:2015tik} including internal symmetries such as color and spin. For simplicity, we begin with a scalar (spin-less) massless particle coupled to a background non-Abelian gauge field, where the one loop-effective action is
\begin{align}\label{eq:effetciveaction1}
\Gamma[A]=-\text{Tr}\,\log(-i D^2[A])\,.
\end{align}
Here $D_\mu = \partial_\mu-igA_\mu$ is the non-Abelian covariant derivative in arbitrary representation and $\text{Tr}$ denotes an infinite dimensional functional trace over spatial coordinates as well as internal symmetries.  Using the heat-kernel representation of the logarithm, Strassler  \cite{Strassler:1992zr} showed that \Eq{eq:effetciveaction1} can be written as a quantum mechanical path integral
\begin{align}\label{eq:effectiveaction2}
\Gamma[A]=\int\limits_0^\infty \frac{dT}{T} \mathcal{N}\, Dx \,\text{tr}\,\mathcal{P}\exp{\left[ i\int\limits_0^T d\tau \left( \frac{\dot{x}^2}{2\epsilon} +g A_\mu[x]\dot{x}^\mu \right)\right]}\,,
\end{align}   
where the functional trace over position is now written as a path integral of a pointlike particle satisfying a trajectory in proper time $\tau$ with the position eigenvalue $x_\mu(\tau)$. The einbein $\epsilon$ is an arbitrary positive parameter that we will discuss further shortly.
\Eq{eq:effectiveaction2} contains the trace in color of the matrix valued Wilson line of the external gauge field along the world-line. 

Remarkably, this proper-time ordered color trace can be expressed as a functional integral just as is the case for the spatial coordinates. In particular, D'Hoker and Gagne derived the following very elegant identity for an $n\times n$ Hermitian matrix $M(\tau)$~\cite{DHoker:1995uyv,DHoker:1995aat}\footnote{A similar representation was suggested previously in \cite{Barducci:1980xk}. Worldline representations for mixed symmetries are discussed in \cite{Edwards:2016acz}.}:
\begin{align}\label{eq:Lagrange}
\text{tr}\, \mathcal{P} \exp{\left[i \int_0^T d\tau M(\tau)\right] } = \int \mathcal{D}\phi\int \mathcal{D}\lambda^\dagger \mathcal{D}\lambda\, e^{i\phi (\lambda^\dagger \lambda +\frac{n}{2}-1)}\nonumber\\
\times\exp{\left[ i \int_0^T d\tau (i \lambda^\dagger \frac{d\lambda}{d\tau} + \lambda^\dagger M \lambda ) \right]}\,,
\end{align}
where $\int \mathcal{D}\phi\equiv  \left( \frac{\pi}{T}\right)^n\sum_\phi$ and  $\lambda$, $\lambda^\dagger$ are Grassmann valued eigenvalues of fermionic creation and annihilation operators. The exponential factor $e^{i\phi (\lambda^\dagger \lambda +\frac{n}{2}-1)}$ is a constraint that restricts the fermion creation and annihilation operators to act on a finite dimensional representation of the internal symmetry group \cite{Barducci:1976xq,JalilianMarian:1999xt}.  

Substituting \Eq{eq:Lagrange} in \Eq{eq:effectiveaction2}, one obtains the Schwinger-Keldysh real-time effective action to be
\begin{align}\label{eq:WLpathintSK}
\Gamma_\mathcal{C}[{A}; \chi]&=\int d^4x_{i}^+d^4x^-_{i}\int d\lambda^+_{i}d\lambda^-_{i} \int d{\lambda^\dagger}^+_{i}d{\lambda^\dagger}^-_{i} \nonumber\\&\times \chi_{A}(x^+_{i},x^-_{i},\lambda^+_{i},\lambda^-_{i}, {\lambda^\dagger}^+_i, {\lambda^\dagger}^-_i)\nonumber\\&\times\int\limits_\mathcal{C}\mathcal{D}\epsilon\mathcal{D}\phi\int\limits_\mathcal{C}\mathcal{D}x\, 
\int\limits_\mathcal{C}\mathcal{D}\lambda^\dagger \mathcal{D}\lambda \; e^{i S_\mathcal{C}[A] }\,.
\end{align}
In this expression, $\chi_A$ is the density matrix at initial time $\tau_0$ with $(x_i^\pm,\lambda_i^\pm,{\lambda^\dagger}_i^\pm)\equiv (x^\pm,\lambda^\pm,{\lambda^\dagger}^\pm)(\tau_0)$, $\pm$ denotes support on the upper/lower Keldysh contour and $\mathcal{C}$ denotes integration along this contour with the (massless) SK action given by
\begin{align}
S_\mathcal{C}[A]=\int\limits_\mathcal{C}d\tau\big\{&\frac{\dot{x}^2}{2\epsilon} + g \dot{x}_\mu \, \lambda^\dagger  A^\mu \lambda + i\lambda^\dagger \dot{\lambda}+ \phi[\lambda^\dagger \lambda + \frac{n}{2}-1]\big\}\,.
\end{align}
In going from \Eq{eq:effectiveaction2} to \Eq{eq:WLpathintSK}, we replaced the $dT/T$ integral  by a more general expression, where the einbein $\epsilon$ is promoted to a dynamical variable and integrated over. The integration over the einbein can be understood as integration over a gauge redundancy, which results from the world line action in \Eq{eq:effectiveaction2} being invariant under rescaling of the world-line parameter $\tau\rightarrow \tau'$. As shown in the Euclidean formulation of \cite{Bastianelli:2006rx}, one can ``BRST-fix" this gauge freedom in \Eq{eq:WLpathintSK} to arrive at  \Eq{eq:effectiveaction2}, where the $dT/T$ integral is a remnant of this construction\footnote{This can be interpreted as the integral over each 
gauge equivalent configuration normalized by the volume of the gauge group (not explicitly written).}. 

The einbein representation in \Eq{eq:WLpathintSK} is advantageous when taking the saddle point of the SK path integral.
To make contact with the phase space formulation of the problem, we can write the action as  $S_\mathcal{C}\equiv \int_\mathcal{C} d\tau (i\lambda^\dagger \cdot\dot{\lambda} + p\cdot \dot{x} -H)$ with the Hamiltonian 
\begin{align}
H=\frac{\epsilon}{2}  P^2- \phi[\lambda^\dagger \lambda + \frac{n}{2}-1]\,,
\end{align}
where $P_\mu = p_\mu - g \lambda^\dagger_a \mathcal{A}_\mu^c t^{c}_{ab} \lambda_b$ is the kinetic momentum, and explicitly write the canonical conjugate momentum in the SK phase space measure.  Employing the truncated Wigner  approximation (also known as the classical-statistical approximation), the Keldysh action can be expanded as \cite{Polkovnikov:2009ys,Jeon:2013zga,Berges:2015kfa}
\begin{align}\label{eq:TWAaction1}
S_\mathcal{C}=S_0+&\int d\tau\, \Big(      \left[\dot{\bar{p}}-\frac{\partial}{\partial \tilde{x}}H\right] \tilde{x} -\left[\dot{\bar{x}}+\frac{\partial}{\partial \tilde{p}}H\right] \tilde{p}\nonumber\\ &+ \left[i\dot{\bar{\lambda}}-\frac{\partial}{\partial \tilde{\lambda}^\dagger} H \right]\tilde{\lambda}^\dagger + \left[i\dot{\bar{\lambda}}^\dagger+\frac{\partial}{\partial \tilde{\lambda}} H \right]\tilde{\lambda}    \Big)+O(\hbar^2)
\end{align}
where we introduced the Keldysh basis $\bar{z}\equiv\frac{1}{2}(z^++z^-)$, $\tilde{z}\equiv z^+-z^-$, with $z$ denoting collectively any of the SK path integral variables ($z^\pm\in \{x^\pm,p^\pm,\lambda^\pm,\lambda^{\dagger\pm}\}$)~\cite{longpaper,Polkovnikov:2009ys,Jeon:2013zga,Berges:2015kfa}. 
Here $\bar{z}$ are interpreted as classical variables while $\tilde{z}$ are the ``quantum" degrees of freedom measured in units of $\hbar$~\cite{Polkovnikov:2009ys,Jeon:2013zga,Berges:2015kfa}.
One obtains $S_0=\int d\tau\, \left\{  -({\tilde{\epsilon}}/{2}) \bar{P}^2  \right\} + \tilde{\phi}(\bar{\lambda}^\dagger\bar{\lambda} + \frac{n}{2}-1 )$
and the phase space measure becomes $\int_\mathcal{C} \mathcal{D}z\equiv \int \mathcal{D}z^+\mathcal{D}z^-= \int\mathcal{D}\bar{z}\mathcal{D}\tilde{z}$. 

To make contact with the phase space formulation of quantum mechanics, we can express the initial density matrix in terms of its Wigner transform  $W_A^{\chi}$ as\footnote{Note that in writing this 
expression, we have used the fact that a fermion coherent state Wigner distribution can be expressed as \cite{Ohnuki:1978jv,Abe:1989mr}, $\chi(\lambda^+,\lambda^-,{\lambda^\dagger}^+,{\lambda^\dagger}^-)  \equiv \int d^4\bar{p}_\lambda d^4\bar{p}_{\lambda^\dagger} W( \bar{\lambda}, \bar{\lambda}^\dagger, \bar{p}_\lambda, \bar{p}_{\lambda^\dagger} ) $ $\times\exp({\tilde{\lambda}\bar{p}_\lambda + \tilde{\lambda}^\dagger \bar{p}_{\lambda^\dagger}})$. Further, noting that $\bar{p}_\lambda=i \bar{\lambda}/2^\dagger$ and $\bar{p}_{\lambda^\dagger}=i \bar{\lambda}/2$ are second class constraints, we obtain the form of the Wigner function in \Eq{eq:Wigner}.} 
\begin{align}\label{eq:Wigner}
\chi_{A}(x^+_{i},&x^-_{i},\lambda^+_{i},\lambda^-_{i}, {\lambda^\dagger}^+_i, {\lambda^\dagger}^-_i) \nonumber\\&\equiv \int \frac{d^4\bar{p}_{i}}{(2\pi)^4} \,
W_A^\chi(\bar{x}_{i},\bar{p}_{i},\bar{\lambda}_{i},\bar{\lambda}_{i}^\dagger) \, e^{ i (\bar{p}_{i}\cdot \tilde{x}_{i}+\frac{1}{2}\bar{\lambda}_{i}^\dagger\cdot \tilde{\lambda}_{i}+\frac{1}{2}{\bar{\lambda}}_{i}\cdot \tilde{\lambda}_{i}^\dagger   ) }\,.
\end{align}
Substituting \Eq{eq:TWAaction1} and \Eq{eq:Wigner} into \Eq{eq:WLpathintSK}, the path integral can be performed,
\begin{align}
\Gamma_\mathcal{C}\approx \int &d^4 \bar{x}_i d^4 \bar{p}_i d\bar{\lambda}_id\bar{\lambda}^\dagger_i \; W_A^\chi(\bar{x}_{i},\bar{p}_{i},\bar{\lambda}_{i},\bar{\lambda}_{i}^\dagger) \nonumber\\ \times \int_\mathcal{C} Dx&  Dp D{\lambda} D{\lambda}^\dagger D{\epsilon}D{\phi}
 \;\exp \Big\{iS_0+ i\int d\tau\, \Big(      \left[\dot{\bar{p}}-\frac{\partial H}{\partial \tilde{x}}\right] \tilde{x} \nonumber\\& -\left[\dot{\bar{x}}+\frac{\partial H}{\partial \tilde{p}}\right] \tilde{p}- \left[i\dot{\bar{\lambda}}-\frac{\partial H}{\partial  \tilde{\lambda}^\dagger} \right]\tilde{\lambda}^\dagger - \left[i\dot{\bar{\lambda}}^\dagger-\frac{\partial H}{\partial \tilde{\lambda}}  \right]\tilde{\lambda}    \Big)\Big\}\,\nonumber\\
 =  \int& d^4 \bar{x}_i d^4 \bar{p}_i d\bar{\lambda}_id\bar{\lambda}^\dagger_i  \; W_A^\chi(\bar{x}_{i},\bar{p}_{i},\bar{\lambda}_{i},\bar{\lambda}^\dagger_i) \int\limits D\bar{x}  D\bar{p} D\bar{\lambda} D\bar{\lambda}^\dagger  \nonumber\\\times &D\bar{\epsilon}D\bar{\phi}\prod\limits_\tau \delta(\dot{\bar{p}}(\tau)-\dot{{p}}_{cl})\,
\delta(\dot{\bar{x}}(\tau)-\dot{{x}}_{cl})\,\delta(\dot{\bar{\lambda}}(\tau)-\dot{{\lambda}}_{cl})\nonumber\\ \times&\delta(\dot{\bar{\lambda}}^\dagger(\tau)-\dot{{\lambda}}^\dagger_{cl})\,\delta(\bar{P^2})\delta(\bar{\lambda}_a^\dagger \bar{\lambda}_a(\tau) + \frac{n}{2}-1)\,.
\label{eq:finalTWA}
\end{align}
The variables $\dot{x}_{cl}, \dot{p}_{cl}, \dot{\lambda}_{cl}, \dot{\lambda}^\dagger_{cl}$ satisfy the classical equations of motion which we will specify shortly in section \ref{sec:color}. 
The last two delta functions in \Eq{eq:finalTWA} impose classical spectral constraints
and are obtained by integration over $\tilde{\epsilon}$ and $\tilde{\phi}$. 

\Eq{eq:finalTWA} contains all the ingredients to construct classical phase space distributions with internal symmetries. Specifically, $W_A^{\chi}(\bar{x}_{i},\bar{p}_{i},\bar{\lambda}_{i},\bar{\lambda}^\dagger_i)$ is a generalized Wigner distribution that coincides with the classical Liouville density in the $\hbar=0$ limit. The Wigner distribution at a given world-line proper time can be 
obtained from its initial distribution at $\tau_0$ from the relation
\begin{align}\label{eq:defW}
W_A^\chi[&\bar{x}(\tau),\bar{p}(\tau),\bar{\lambda}(\tau),\bar{\lambda}^\dagger(\tau)  ] \nonumber\\\equiv  &\int d^4 \bar{x}_i d^4 \bar{p}_i d\bar{\lambda}_id\bar{\lambda}^\dagger_i \; W_A^{\chi}[\bar{x}_{i},\bar{p}_{i},\bar{\lambda}_{i},\bar{\lambda}_{i}^\dagger)
]\nonumber\\&\times\int\limits D\bar{x}'  D\bar{p}' D\bar{\lambda}' {D\bar{\lambda}^\dagger}'   D\bar{\epsilon}'D\bar{\phi}'\nonumber\\ \times&\prod\limits_{\tau'\le \tau} \delta(\dot{\bar{p}}(\tau')-\dot{{p}}_{cl})\,
\delta(\dot{\bar{x}}(\tau')-\dot{{x}}_{cl})\,\delta(\dot{\bar{\lambda}}(\tau')-\dot{{\lambda}}_{cl})\nonumber\\ \times&\delta(\dot{\bar{\lambda}}^\dagger(\tau')-\dot{{\lambda}}^\dagger_{cl})\,\delta(\bar{P^2(\tau')})\delta(\bar{\lambda}_a^\dagger \bar{\lambda}_a(\tau') + \frac{n}{2}-1)\,,
\end{align}
where $\mathcal{D}\bar{z}\equiv \prod_{\tau' < \tau} d\bar{z}(\tau')$\footnote{We emphasize that the delta functions in \Eq{eq:defW} include the last $\tau$ slice, while the path integral excludes it.}. This relation can equivalently be expressed as the solution to Liouville's equation,
\begin{align}\label{eq:Liouville1}
\frac{d}{d\tau} W_A^\chi=\left(\dot{\bar{x}}_\mu\frac{\partial}{\partial \bar{x}_\mu}+\dot{\bar{P}}_\mu\frac{\partial}{\partial \bar{P}_\mu} +\dot{\bar{\lambda}}_a\frac{{\partial}}{\partial \bar{\lambda}_a}+\dot{\bar{\lambda}}^\dagger_a\frac{\partial}{\partial \bar{\lambda}^\dagger_a} \right)W_A^\chi\,.
\end{align}
for a given initial condition $W_A^{\chi}[\bar{x}_{i},\bar{p}_{i},\bar{\lambda}_{i},\bar{\lambda}_{i}^\dagger ]$ at $\tau=\tau_0$. In its most general relativistic formulation, Liouville's equation satisfies $d W_A^{\chi}/d\tau=0$, which represents the incompressibility of the semi-classical phase space distribution. Note that the quantum nature of $W_A^\chi$ is embedded in the stochastic distribution of initial conditions; the subsequent evolution of these initial configurations is classical. 

In appendix \ref{app:phasepsace}, we discuss the equivalence of the Wigner distribution in TWA  (\Eq{eq:defW}) and Liouville's equation (\Eq{eq:Liouville1}). In the semi-classical limit, phase space is incompressible and is a simple consequence of the fact that the variables in our construction are canonical\footnote{The reverse is not necessarily true.}. We will return to the topic of compressibility in section \ref{sec:anomaly}.

As the subscript $A$ denotes, the Wigner distribution in \Eq{eq:defW} is further embedded in a larger path integral of dynamical gauge fields~\cite{Mueller:2017lzw,Mueller:2017arw}. The resulting quantum kinetic theory thus includes a double average, where one splits fluctuating modes, according to their lifetimes, into quasi-static classical fields and shorter lived dynamical modes in the presence of these fields. Two examples of phenomena in QCD where such a separation of scales is employed are finite temperature QCD plasmas~\cite{Braaten:1989mz,Braaten:1995cm,Bodeker:1998hm,Litim:2001db,Litim:1999ns,Litim:1999id,Blaizot:2001nr,Litim:2001je} and small x physics \cite{McLerran:1993ni,Gelis:2010nm,Blaizot:2016qgz}. 

\Eqs{eq:finalTWA}{eq:Liouville1} specify the semiclassical limit of color through the ``classical" Grassmann variables $\lambda,\lambda^\dagger$; as we shall shortly discuss in section \ref{sec:spin}, a similar result is obtained when one includes spin through the anticommuting coordinates $\psi^\mu$. 
%
%
%
\section{Phase space representations of color}\label{sec:color}
In this section, we will discuss further the properties of the Wigner distribution\footnote{For simplicity, we will here, and henceforth, drop the bar symbol representing classical phase space coordinates.} $W_A^\chi(x,P,\lambda,\lambda^\dagger)$ which obeys the Liouville equation defined in \Eq{eq:Liouville1}. We will show that this Wigner distribution can equivalently be reexpressed in  terms of the color charges $Q^a$ that are constructed from Grassmann bilinears. As we will discuss, this formulation is elegant and potentially powerful. 


In \Eq{eq:Liouville1}, the Euler-Lagrange equations of motion derived from the world-line effective action are \cite{Balachandran:1977ub,Barducci:1976xq,JalilianMarian:1999xt}
\begin{align}
\dot{x}^\mu &=  v^\mu\,,\label{eq:EOM1}\\
\dot{P}^\mu &= g  \lambda^\dagger_a F^{c,\mu\nu} t^c_{ab} \lambda_b v_\nu\,,\\
\dot{\lambda}^\dagger_a &= -ig \,   v^\mu t^c_{ab} A_{\mu }^c \, \lambda^\dagger_b\,,\\
\dot{\lambda}_a &= ig \,   v^\mu t^c_{ab} A_{\mu }^c \lambda_b\,,
\label{eq:EOM3}\
\end{align}
where $v^\mu\equiv \epsilon P^\mu=\epsilon[p_\mu-g A^a_\mu(x)Q^a]$.  
 The corresponding Dirac brackets are defined as
\begin{align}
\{A,B\}=&A\left( \frac{\overleftarrow{\partial}}{\partial x^\mu} \frac{\overrightarrow{\partial}}{{\partial} P^\mu}-\frac{\overleftarrow{\partial}}{\partial P^\mu} \frac{\overrightarrow{\partial}}{\partial x^\mu} \right)B\nonumber\\&
-i A\left( \frac{\overleftarrow{\partial}}{\partial {\lambda^\dagger}_a } \frac{\overrightarrow{\partial}}{\partial \lambda_a}+\frac{\overleftarrow{\partial}}{\partial \lambda_a } \frac{\overrightarrow{\partial}}{ \partial {\lambda^\dagger}_a}\right)B\,,
\label{eq:Dirac}
\end{align}
and give 
\begin{align}
\{x^\mu, P^\nu \}&=g^{\mu\nu}\,,\\
\{ P^\mu,P^\nu \} &= g \lambda^\dagger F^{\mu\nu}\lambda \,,\\
\{\lambda^\dagger_a,\lambda_b \} &= -i \delta_{ab}\,,\label{eq:grassmannPoisson}\\
\{ P_\mu, F_{\alpha\beta}\} &= -(D_\mu F_{\alpha\beta})\,.
\end{align}
The classical color phase space measure at fixed proper time $\tau$ follows directly from the saddle point limit of the SK path integral,  \Eq{eq:defW}: 
\begin{align}\label{eq:definition}
\int d\lambda(\tau)\,d\lambda^\dagger(\tau) \equiv \int d^n\lambda(\tau)\, d^n\lambda^\dagger(\tau)\,\delta(\lambda_a^\dagger (\tau)\lambda_a(\tau) +\frac{n}{2}-1)  \,,
\end{align}
where the integration is over the ``classical" Keldysh coordinates which we denoted with bars in \Eq{eq:finalTWA}.  The phase space measure represents the integration over the phase space variables for one specific slice in proper time; from now on, we will drop the explicit label $\tau$. Also, $n$ is the dimension of the representation of the color group; we will restrict ourselves to $SU(3)$ in this work. 

The Liouville density $W_A^\chi(x,P,\lambda,\lambda^\dagger)$ can be expanded as a power series in $\lambda,\lambda^\dagger$, and is Grassmann even and real. It can therefore be parametrized by the bilinears $\lambda^\dagger_a \Gamma_{ab}\lambda_b$, where $\Gamma$ are Hermitian matrices. 
A natural choice for $\Gamma$ are the eight $SU(3)$ generators and the identity, which span a complete trace-orthonormal set,
\begin{align}\label{eq:defQ}
Q^a \equiv \lambda_c^\dagger t^a_{cd} \lambda_d\,,
\end{align}
where insertion of the identity  $\lambda_a^\dagger \delta_{ab} \lambda_b$ produces a phase space invariant and can be dropped.
The classical Dirac brackets of the bilinears follow from \Eq{eq:grassmannPoisson},
\begin{align} \label{eq:coloralgebra}
\{Q^a,Q^b\} = \lambda^\dagger [t^a,t^b]\lambda = i f^{abc}Q^c\,,
\end{align}
realizing the $SU(N_c)$ group algebra
\footnote{The ``classical" color realization in \Eq{eq:coloralgebra} thereby does not imply a large color representation via a large number of color sources \cite{Jeon:2004rk}, but is valid for any representation of $SU(3)$.}.
\Eqs{eq:EOM1}{eq:EOM3}, along with \Eq{eq:defQ}, reproduce Wong's equations~\cite{Wong:1970fu} for the precession of color charges in a non-Abelian background field, thereby providing a first principles derivation for the same \cite{Balachandran:1977ub,Barducci:1976xq}, but also cleanly establishing its provenance and regime of applicability~\cite{JalilianMarian:1999xt} in QCD. 

The phase space measure of the $Q^a$ charges was not discussed in these papers. However, by simply defining 
\begin{align}
\int dQ  \equiv \int d\lambda\,d\lambda^\dagger\,,
\end{align}
the following identities are obtained:
\begin{align}
\int dQ &= 0\,,\label{eq:identity-a}\\
\int dQ \,Q^a &= 0\,,\label{eq:identity-b}\\
\int dQ\, Q^a Q^b &= \frac{1}{2}\delta^{ab}\,,\label{eq:ident1}\\
\int dQ\, Q^a Q^b Q^c&=\frac{A_R}{2}d^{abc}\,,\label{eq:ident2}
\end{align}
where $A_R$ is the so-called anomaly coefficient of the representation \cite{Peskin:1995ev}. Integrals of higher polynomials of $Q$'s vanish by Grassmann nilpotency. (\Eq{eq:ident1} and \Eq{eq:ident2} are proven in appendix \ref{app:color}.)

The most general form of the phase space distribution is thus
\begin{align}\label{eq:generaldist}
W_A^\chi(x,P,\lambda&,\lambda^\dagger)\rightarrow f_A(x,P,Q) = c_0 f(x,P) + c_1f^a(x,P)Q^a\nonumber\\&+ c_2f^{ab}(x,P)Q^aQ^b+ c_3f^{abc}(x,P)Q^aQ^bQ^c \,,
\end{align}
with real coefficients $c_i$, which can be determined by taking moments,
\begin{align}
f(x,P)&\equiv \int dQ\, f(x,P,Q)\,,\label{eq:mom0}\\
f^a(x,P)&\equiv \int dQ\, Q^a \,f(x,P,Q)\,,\\
f^{ab}(x,P)&\equiv \int dQ\, Q^a Q^b  f(x,P,Q)\,,\\
f^{abc}(x,P)&\equiv \int dQ\, Q^a Q^b Q^c f(x,P,Q)\,.
\label{eq:mom3}\,
\end{align}
Using the properties of the Grassmann variables, one obtains the form of the phase space distribution function to be 
\begin{align}\label{eq:generaldist}
f(&x,P,Q) \nonumber \\ &=f(x,P)\big[ 1 + \frac{2}{A_Rd^2}d^{abc}Q^a Q^b Q^c\big] + 2 f^a(x,P) Q^a\,,
\end{align}
where $d^2\equiv d^{abc}d^{abc}=N_c^2-4$. 

\Eq{eq:generaldist} is a key result of our study and is a direct consequence of the Grassmann color algebra obtained from the saddle point limit of the  SK world-line path integral.
Details of this derivation can be found in appendix \ref{app:color}, where we also show that higher color moments of the phase space distribution are completely determined by the lower singlet and octet moments:
\begin{align}
f^{ab}(x,P)&=\frac{1}{2}\delta^{ab} \, f(x,P)+ A_R\, d^{abc} f^c(x,P)\,,\\
f^{abc}(x,P)& = \frac{A_R}{2} d^{abc}\,f(x,P)\,.\label{eq:rel2}
\end{align}
Similar phase space representations were discussed previously~\cite{Heinz:1983nx,Heinz:1984yq,Heinz:1985qe,Kelly:1994ig,Kelly:1994dh,Litim:2001db,Litim:1999ns,Litim:1999id}. However the underlying Grassmann algebra was not considered in these works; in its absence, one may conclude that an infinite tower of independent moments analogous to \Eqs{eq:mom0}{eq:mom3}\ is feasible. In contrast, we have shown 
that the entire content of the phase space dynamics is captured by two moments giving the simple form in \Eq{eq:rel2}. Our construction is therefore potentially very useful in further 
development of the transport theory of color charges. 

The bilinears $Q^a(\tau)$ parametrize orbits of the gauge group $SU(3)$ in the classical limit. They coincide
with the  commuting Darboux variables discussed by Alekseev, Faddeev and Shatashvili \cite{Alekseev:1988vx} (AFS) -- albeit their origin in terms of Grassmann coherent states was not considered there. In that work, AFS used spherical coordinates to derive an action principle parametrizing orbits of various compact Lie groups. Specifically, for $SU(2)$, they specified the Darboux variables by the spherical coordinates $\phi,\theta$ and imposed the quantization constraint $\sqrt{Q^aQ^a}=m\sim \hbar$. As a result, the authors arrived at the topological action
\begin{align}
S=m\int \cos\theta\,d\phi + \gamma \int d\phi\,,
\end{align}
containing a Berry monopole at $Q^aQ^a=m^2=0$. Here $\gamma=1/2$ if $m$ is half-integer and $\gamma=0$ if $m$ is integer.  This construction is for fixed spin length $m\sim\hbar$ and it is unclear what happens for $\hbar\rightarrow 0$.
In contrast, no topological term arises in our world-line construction. The classical limit $\hbar\rightarrow 0$ is understood to be the saddle point approximation to the SK world-line path integral, where classical and quantum degrees of freedom are manifest in the Keldysh basis \cite{Polkovnikov:2009ys,Jeon:2013zga,Berges:2015kfa}.%
%
%
%
%
%
\section{Representations of spin and chirality}\label{sec:spin}
We shall now extend the construction developed in the previous section to include spin, with a focus on relativistic spin-$1/2$ fermions. The Grassmann representation of phase space was obtained  by Berezin and Marinov over 40 years ago~\cite{Berezin:1976eg}. The Hamiltonian for a massive colored spin-$1/2$ fermion is\footnote{Hamiltonians for spin-1 can be found in \cite{ Ahmadiniaz:2018olx}.} 
\begin{align}
{\tilde H}= &\frac{\epsilon}{2}\left( P^2 + m^2 +ig\psi^\mu \lambda^{\dagger b}F^a_{\mu\nu}(x)t^a_{bc}\, \lambda^c\psi^\nu \right)\nonumber\\&+\frac{i}{2}\left( P_\mu \psi^\mu +m\psi_5  \right)\chi 
\label{eq:Hamiltspinandcolor}\,,
\end{align}
where $g$ is the QCD coupling,  $\psi^\mu,\psi_5$ are Grassmann spin variables and $\chi$ is a Grassmann valued supersymmetric partner of the einbein parameter $\epsilon$. 
Using this expression for the Hamiltonian and \Eq{eq:Dirac}, the classical equations of motion are 
\begin{align}
\dot{x}^\mu &= \epsilon P^\mu\,,\label{eq:emospin1}\\
\dot{P}^\mu &= \epsilon g F^{a,\mu\nu} Q^a P_\nu  -\frac{i\epsilon g}{2} \psi^\alpha (D^\mu F_{\alpha\beta})^a Q^a \psi^\beta\,,\\
\dot{\psi}^\mu &=\epsilon g   F^{a,\mu\nu} Q^a \, \psi_\nu \,,\label{eq:emospin3}\\
\dot{\lambda}_a^\dagger&=-igv^\mu t^{c}_{ab} A_\mu^c \lambda_b^\dagger - \frac{\epsilon g}{2} \psi^\mu F_{\mu\nu}^b t^b_{ac}\lambda^\dagger_c \psi^\nu\,,\label{eq:emospin4}\\
\dot{\lambda}_a&=igv^\mu t^{c}_{ab} A_\mu^c \lambda_b^\dagger + \frac{\epsilon g}{2} \psi^\mu F_{\mu\nu}^b t^b_{ac}\lambda_c \psi^\nu\,,\label{eq:emospin5}
\end{align}
where $v^\mu\equiv \epsilon P^\mu$ and $Q^a$ as in \Eq{eq:defQ}. For a suitable choice of $\chi$, $\psi_5$ is not dynamical and can be dropped from the equations of motion. The Grassmann coordinates obey 
$\{ \psi^\mu, \psi^\nu\}=-ig^{\mu\nu}$.  

The most general form of the Liouville distribution with spin written down by Berezin and Marinov \cite{Berezin:1976eg} -- now extended to color -- is, 
\begin{align}\label{eq:Liouville} 
W_A^\chi(x,P,&\lambda,\lambda^\dagger,\psi)= W_A^\chi(x,P,\lambda,\lambda^\dagger)\Big[ \Sigma_\mu(x,P,\lambda,\lambda^\dagger)\nonumber\\&\times v_\lambda\, \psi^\mu  \psi^\lambda - \frac{i}{6}\epsilon_{\mu\nu\alpha\beta} v^\mu v_\lambda \psi^\nu\psi^\alpha \psi^\beta\psi^\lambda \Big]\,,
\end{align}
which is uniquely parametrized by a pseudo-vector $\Sigma_\mu(x,P,\lambda,\lambda^\dagger)$. As shown in \cite{Berezin:1976eg}, this form of the distribution function equivalently imposes the spectral constraint\footnote{As such, it might be understood as a classical equivalent of the GSO projection \cite{Gliozzi:1976qd,Gliozzi:1976jf} in fermionic extensions of bosonic string theory \cite{Ramond:1971gb,Neveu:1971fz,Gervais:1971ji}, removing unphysical degrees of freedom from the classical equations of motion.} given by the second term of \Eq{eq:Hamiltspinandcolor}. 
As in the case of color discussed in the previous section, we can organize the Grassmann spin variables into the bilinears
\begin{align}\label{eq:spinbilinear}
S_{\mu\nu}=-i\psi_\mu\psi_\nu\,,
\end{align}
which can straightforwardly shown to satisfy the Poisson bracket relation
\begin{align}
\{ S_{\mu\nu}, S_{\alpha\beta} \} &= -g_{\alpha\mu} S_{\beta \nu} + g_{\beta\mu}S_{\alpha\nu} - g_{\alpha\nu}S_{\mu\beta}+g_{\beta\nu}S_{\mu\alpha}\,.
\end{align}
The Liouville density can likewise, through a canonical transformation, be replaced by
\begin{align}
W_A^\chi(x,P,\lambda,\lambda^\dagger,\psi)\rightarrow f_A(x,P,Q,S)\,,
\end{align}
where
\begin{align}
\label{eq:LiouvilleS}
&f_A(x,P,Q,S)\nonumber\\&= f_A(x,P,Q)\left[ i \Sigma_\mu(x,P,Q) S^{\mu\nu} v_\nu - \frac{i}{6}\epsilon_{\mu\nu\alpha\beta} v^\mu S^{\nu\alpha} S^{\beta\lambda} v_\lambda \right]\,.
\end{align}
Further, the  Grassmann spin phase space measure, in analogy to \Eq{eq:definition} can be represented as 
\begin{align}
 \int dS\equiv -i\int d\psi_0 d\psi_1 d\psi_2 d\psi_3 \,,
\end{align}
which yields the identities
\begin{align}
\int dS &= 0\,,\label{eq:identspin1}\\
\int dS \, S_{\mu\nu}&= 0\,,\\
\int dS \, S_{\mu\nu}S_{\alpha\beta}&= i\epsilon_{\mu\nu\alpha\beta}\,.\label{eq:identspin3}
\end{align}

The pseudo-vector $\Sigma_\mu$ defined in \Eq{eq:Liouville} can equivalently be defined as the first moment of the phase space distribution:
\begin{align}
\Sigma_\mu(x,P,Q)\equiv \int dS\, \sigma_\mu(P,S) \,f(x,P,Q,S)\,,
\end{align} 
where we abbreviated 
\begin{align}\label{eq:spinvariable}
\sigma_\mu(P,S) \equiv\frac{1}{2}\epsilon_{\mu\nu\alpha\beta} v^\nu S^{\alpha\beta}\,.
\end{align}
Using \Eq{eq:LiouvilleS} and \Eqs{eq:identspin1}{eq:identspin3}
\begin{align}
\dot{\sigma}_\mu=&\{ \sigma_\mu,{\tilde H}\} = \frac{g}{P^0}F^a_{\mu\nu}Q^a\,\sigma^\nu + \frac{2g}{P^0} \sigma_\alpha F^{a,\alpha\beta}Q^a v_\beta\, v_\mu \nonumber\\&+ \frac{g}{2(P^0)^2} \big[ \sigma_\nu (D^{\nu} \tilde{F}^{\alpha\beta})^a Q^a\, \sigma_\alpha v_\beta \,v_\mu \nonumber\\&\qquad\qquad-
 v_\nu (D^{\nu} \tilde{F}^{\alpha\beta})^a Q^a\, \sigma_\alpha v_\beta \,\sigma_\mu\big]\,,
\label{eq:spin-precession}
\end{align}
where $\tilde{F}^{a,\mu\nu}=\epsilon^{\mu\nu\alpha\beta}F_{\alpha\beta}^a/2$. Note that this is the covariant generalization of the spin precession of the  quantum mechanical spin three vector 
defined to be $\sigma^i = -\frac{i}{2}\epsilon^{ijk}\psi^j \psi^k$.  

Our derivation also allows us to identify  $\Sigma^\mu(x,P,Q)$ as the proper definition of the 
Pauli-Lubanski spin pseudo-vector in quantum field theory\footnote{We remark that $\sigma_\mu$ defined in \Eq{eq:spinvariable} is equivalent to the phase space coordinates suggested in 
\cite{Heinz:1983nx,Heinz:1984yq,Heinz:1985qe}. This identification however did not take into account the Grassmann origin of $S_{\mu\nu}$. More importantly,  $\sigma_\mu$ is not a canonical variable; replacing $S_{\mu\nu}$ by $\sigma_\mu$ fails to preserve the Poisson bracket algebra satisfied by the former.}.
Relative to  \Eq{eq:spin-precession}, the time evolution of $\Sigma_\mu$ has the simpler form
\begin{align}\label{eq:BMT}
\dot{\Sigma}_\mu(x,P,Q)= &\frac{g}{P^0}F^a_{\mu\nu}Q^a\,\Sigma^\nu(x,P,Q) \nonumber\\&+ \frac{2g}{P^0} \Sigma_\alpha(x,P,Q) F^{a,\alpha\beta}Q^a v_\beta\, v_\mu\,.
\end{align}
\Eq{eq:BMT} is the Bargmann-Michel-Telegdi (BMT) equation~\cite{Bargmann:1959gz} mentioned previously. It describes the precession of the spin pseudo-vector generalized to relevant inhomogeneous QCD color backgrounds.  This is a physical quantity;  the relativistic  BMT equation (in QED) is routinely used in high energy accelerator physics to describe spin precession of relativistic spinning electrons and protons in external electromagnetic fields~\cite{Derbenev:1973ia,Chao:1980fz,Montague:1983yi,Bunce:2000uv}. Similarly, from \Eqs{eq:emospin4}{eq:emospin5}, we obtain Wong's equation for the color precession of spin-$1/2$ fermions in background fields\footnote{Solutions to Wong's equations without spin are discussed in \cite{Kosyakov:1998qi}.},
\begin{align}\label{eq:Wong}
\dot{Q}^a=-igv^\mu f^{abc} A_\mu^b Q^c - \frac{g \epsilon}{2} f^{abc} \psi^\mu F_{\mu\nu}^b \psi^\nu Q^c\,.
\end{align}

To summarize our discussion, we have obtained a novel expression for the Liouville phase space distribution (given by $f_A(x,P,Q,S)$ in \Eq{eq:LiouvilleS}) and the corresponding phase space measures (given by \Eq{eq:identity-a}-\Eq{eq:ident2} and \Eq{eq:identspin1}-\Eq{eq:identspin3}) that describes the phase space dynamics of  particles with both color and spin internal degrees of freedom. To do so, we identified from first principles in quantum field theory, the canonical Grassmann bilinear variables $Q^a$  and $S^{\mu\nu}$ that satisfy the Wong equation in \Eq{eq:Wong} and the covariant BMT equation in \Eq{eq:BMT}.  We note that besides providing a compact and transparent formulation of semi-classical transport phenomena in quantum field theory, our world-line construction allows one in principle to devise a systematic route beyond the truncated Wigner/classical-statistical semi-classical approximation employed here. 
%
%
%
%
\section{Semi-classical currents and the chiral anomaly}
\label{sec:anomaly}
A transport problem of great interest in a wide range of fields -- from astrophysical phenomena to heavy-ion collisions to strongly correlated electronic materials -- is chiral kinetic theory
~\cite{Son:2012wh, Stephanov:2012ki,Son:2012zy,Chen:2012ca,Gao:2012ix, Chen:2013iga,Chen:2014cla,Basar:2013qia,Stone:2013sga,Dwivedi:2013dea,Stone:2014fja,
Manuel:2014dza,Manuel:2015zpa,Yamamoto:2015gzz,Sun:2016nig,Hidaka:2016yjf,Yamamoto:2017uul,Gao:2017gfq,Pu:2017apt,Gao:2018wmr,Huang:2018wdl}. Here the challenge 
is to find semi-classical phase space distributions for chiral fermions and to account for the chiral anomaly, when defining vector and axial-vector currents. This is also important
in first principles constructions of anomalous hydrodynamics in a number of different physical contexts~\cite{Son:2009tf,Sadofyev:2010pr,Sadofyev:2010is,Nair:2011mk,Hongo:2013cqa,Ng:2014sqa,Karabali:2014vla,Hirono:2014oda,Glorioso:2017lcn}.

Let's consider first in this chiral kinetic theory context the semi-classical vector and axial vector currents defined naively as phase space averages in our generalized phase space framework. To simplify our discussion,
we shall consider only the Abelian case, where the currents are obtained from the phase space average over $\partial S/i\partial A_\mu$,
\begin{align}
\langle J_{L/R}^\mu (x) \rangle \equiv e \int d^4P \, dS  \,\epsilon \left[P^\mu + S^{\mu\nu}\partial_\nu \right]f(x,P,S)\,.
\end{align}
Here, $e$ is the electromagnetic coupling. Using the Abelian equivalent of \Eq{eq:LiouvilleS} and the fact that semi-classically $\Sigma^\mu_{L/R} = \pm P^\mu/2P^0$ for left and right handed particles in the chiral limit \cite{Bargmann:1948ck},
one can explicitly perform the Grassmann integral over $S$ to obtain
\begin{align}
\langle J^\mu(x)\rangle &= \langle J_R^\mu(x)\rangle+\langle J_L^\mu(x)\rangle=  e \int d^4P\,  \epsilon P^\mu f(x,P)\,,\\
\langle J^\mu_5(x)\rangle &= \langle J_R^\mu(x)\rangle-\langle J_L^\mu(x)\rangle \nonumber\\&=e \int d^4P\,  \epsilon\, \epsilon^{\mu\nu\alpha\beta}  P_\beta \partial_\nu[ \Sigma_\alpha(x,P) f(x,P)]\,.
\label{eq:Abelian-currents}
\end{align}
Both currents are classically conserved, as $\partial_\mu \langle J^\mu(x)\rangle=0$ follows directly from Liouville's equation and
$\partial_\mu \langle J^\mu_5(x)\rangle=0$ from the antisymmetry of $\epsilon^{\mu\nu\alpha\beta}$. This straightforward implementation
clearly misses the effects of the anomaly.

The missing piece is obtained from a more careful derivation within the world-line formalism. We will sketch below a variational approach to the derivation of  the semi-classical limit of the world-line representation for chiral fermions by carefully repeating the derivation of section \ref{sec:Grassman}, this time for chiral fermions.
%
%
The corresponding Schwinger-Keldysh real-time many-body world-line path integral is 
\begin{align}\label{eq:effectiveactiongeneral}
\Gamma[A,B]\equiv \text{tr} \int d^4&x_i^+ d^4x_i^- d^4\psi_i^+ d^4\psi_i^- \, \zeta^{A,B}(x_i^+,x_i^-,\psi_i^+,\psi_i^-)\nonumber\\
&\times \int\limits_{x_i^+ }^{x_i^-}\mathcal{D}x \mathcal{D}p \int\limits_{\psi_i^+ }^{\psi_i^-}\mathcal{D}\psi \int  {\mathcal{D}\epsilon\mathcal{D}\chi}e^{iS_\mathcal{C}[A,B]}\,,
\end{align}
where the trace is over  left and right handed chiral sectors and the initial density matrix $\zeta^{A,B}$ is in $2\times 2$ matrix form, respectively \cite{Mueller:2017lzw,Mueller:2017arw}.
This path integral is embedded in a larger path integral including dynamical gauge fields $A_\mu$ and a novel nondynamical variational axial-vector gauge field $B_\mu$ that will be put to zero at the end of the derivation.  The exponential in \Eq{eq:effectiveactiongeneral} can be written as 
\begin{align}
e^{iS_\mathcal{C}[A,B]} \equiv 
 \begin{pmatrix}
  e^{iS_\mathcal{C}[A+B]}  & 0\\
  0 &  e^{iS_\mathcal{C}[A-B]} 
\end{pmatrix}\,,
\end{align}
with the spinning particle action given by
\begin{align}
S[A\pm B]=\int d\tau_\mathcal{C} \left[ p_\mu \dot{x}^\mu + \frac{i}{2}\psi_\mu \dot{\psi}^\mu - H[A\pm B] \right]\,,
\end{align}
where $\mathcal{C}$ denotes the 'in-in' (closed) Keldysh time contour. The generalization of the Hamiltonian in \Eq{eq:Hamiltspinandcolor} is given by 
\begin{align}\label{eq:Hamiltonian}
H[A\pm B]\equiv& \frac{\epsilon}{2}\left(P^2 + i e\psi^\mu F_{\mu\nu}[A\pm B]\psi^\nu \right) \nonumber\\ +& \frac{i}{4}\left( P_\mu \psi^\mu \pm \frac{i}{3} \epsilon_{\mu\nu\alpha\beta} P^\mu \psi^\nu \psi^\alpha \psi^\beta  \right) \chi\,,
\end{align}
with kinetic momentum $P_\mu \equiv p_\mu - e(A\pm B)_\mu$. The anticommuting Lagrange multiplier $\chi$ ensures that the Weyl spectral condition is independently satisfied for both left and right handed fermions~\cite{Barducci:1980wt,Mueller:2017arw}\footnote{We note that 
Berezin and Marinov's phase space distribution, \Eq{eq:Liouville}, reflects precisely the same Weyl spectral constraint as in \Eq{eq:Hamiltonian} in the chiral limit where $\Sigma_\mu\sim \pm P_\mu$.}. 

It is sufficient to expand the effective action to linear order in the variational parameter $B_\mu(y)$, 
\begin{align}\label{eq:expansion}
\Gamma[A,B] = \Gamma[A] + \int d^4y\, \frac{\delta \Gamma[A,B] }{\delta B_\mu(y)}\Big|_{B=0} B_\mu(y)\,,
\end{align}
where the linear term $ \delta \Gamma / i\delta B_\mu(y)\equiv \langle J_{5,\mu}(y)\rangle$ is the expectation value of the axial vector current. 
The world-line representation for the lowest order term $\Gamma[A]$ is 
\begin{align}\label{eq:zerothorder}
\Gamma[A]=\text{tr} \int d^4&x_i^+ d^4x_i^- d^4\psi_i^+ d^4\psi_i^- \, \zeta^A(x_i^+,x_i^-,\psi_i^+,\psi_i^-)\nonumber\\
&\times \int\limits_{x_i^+ }^{x_i^-}\mathcal{D}x \mathcal{D}p \int\limits_{\psi_i^+ }^{\psi_i^-}\mathcal{D}\psi   {\mathcal{D}\epsilon\mathcal{D}\chi}e^{iS_\mathcal{C}[A]}\,,
\end{align}
which is equivalent to \Eq{eq:WLpathintSK} if we replace color by spin. The linear term in \Eq{eq:expansion} can then be written as
\begin{align}\label{eq:linearterm}
\frac{\delta \Gamma[A,B] }{\delta B_\mu(y)}&\Big|_{B=0} = \text{tr} \int d^4x_i^+ d^4x_i^- d^4\psi_i^+ d^4\psi_i^- \,\nonumber\\\times&\Bigg[ \zeta^{A,B}(x_i^+,x_i^-,\psi_i^+,\psi_i^-)\int\limits_{x_i^+ }^{x_i^-} \mathcal{D}x \mathcal{D}p \int\limits_{\psi_i^+ }^{\psi_i^-}\mathcal{D}\psi   \int{\mathcal{D}\epsilon
\mathcal{D}\chi}\,\nonumber\\  &\times\frac{i\delta S_\mathcal{C}[A,B]}{\delta B_\mu(y)}e^{iS_\mathcal{C}[A]}
+ \frac{\delta \zeta^{A,B}(x_i^+,x_i^-,\psi_i^+,\psi_i^-)}{\delta B_\mu(y)} \nonumber\\&\times  \int\limits_{x_i^+ }^{x_i^-}\mathcal{D}x \mathcal{D}p \int\limits_{\psi_i^+ }^{\psi_i^-}\mathcal{D}\psi   \int{\mathcal{D}\epsilon
\mathcal{D}\chi}\, e^{iS_\mathcal{C}[A,B]}\Bigg]\Big|_{B=0}\,.
\end{align}
The first term of \Eq{eq:linearterm} in the square brackets is the expectation value of the axial-vector current given previously in \Eq{eq:currentexpectation},
\begin{align}\label{eq:currentexpectation}
\langle J^\mu_5(y)\rangle =\,\text{tr} \int d^4x_i^+ d^4x_i^- d^4\psi_i^+ d^4\psi_i^-\zeta^A(x_i^+,x_i^-,\psi_i^+,\psi_i^-)\nonumber\\ \times \int\limits_{x_i^+ }^{x_i^-} \mathcal{D}x \mathcal{D}p \int\limits_{\psi_i^+ }^{\psi_i^-}\mathcal{D}\psi \int  {\mathcal{D}\epsilon
\mathcal{D}\chi}
 \begin{pmatrix} J_R^\mu & 0 \\
0 & J^\mu_L
\end{pmatrix}
 e^{iS_\mathcal{C}[A]}\,,
\end{align}
where 
\begin{align}
J_{L/R}^\mu &\equiv \frac{\delta S_\mathcal{C}[A\pm B]}{i\delta B_\mu(y)}\Big|_{B=0}\nonumber\\&= \pm e \int d\tau_\mathcal{C}\, \epsilon\left(  P^\mu - i\psi^\mu\psi^\nu\partial_\nu  \right)\delta[x-y]\,.
\end{align}

To compute the second term of \Eq{eq:linearterm}, we split the initial density matrix,
\begin{align}
\zeta\equiv \zeta^{(0)}+\zeta^{(1)} \,,
\end{align}
where $\zeta^{(0)}$ parametrizes arbitrary occupation numbers of left and right handed fermions at initial time and is independent of $B_\mu$:
\begin{align}
{\zeta}^{(0)}\equiv \begin{pmatrix}
\zeta_R^{A} [x_i^+,x_i^-,\psi_i^+,\psi_i^- ] & 0 \\
0 & \zeta_L^{A} [x_i^+,x_i^-,\psi_i^+,\psi_i^- ]
\end{pmatrix}\,.
\end{align}
Only the second piece $\zeta^{(1)}$ is $B_\mu$-dependent and therefore contributes to the second term in \Eq{eq:linearterm}. It was computed previously in \cite{Mueller:2017arw} to be 
\begin{align}\label{eq:densityvacuum}
\zeta^{(1)} \equiv 2\, \mathbb{I}_{2\times 2} \,\big[ \partial_\mu &B_\mu (\bar{x}_i) - \{ \partial_\mu, B_\nu (\bar{x}_i) \}\bar{\psi}^\nu \bar{\psi}^\mu  \big]\nonumber\\&\times \delta(x^+_i -x^-_i)\, \delta(\psi^+_i -\psi^-_i) \,,
\end{align}
where $\bar{x}_i = (x_i^++x_i^-)/2$ and $\bar{\psi}_i = (\psi_i^++\psi_i^-)/2$. This term can be  interpreted as a vacuum contribution to the initial density matrix and survives as an 
anomalous contribution to $\langle \partial_\mu J_5^\mu\rangle$ when we set $B_\mu\rightarrow 0$, thereby modifying the expectation value of the axial vector current in \Eq{eq:currentexpectation}. 
 
In \cite{Mueller:2017arw}, we showed by analytic continuation in Euclidean space that the anomalous non-conservation of the axial-vector current arises from this contribution and yields the well known expression
\begin{align}\label{eq:currentenonconservation}
\langle \partial_\mu J^\mu_5(y) \rangle = -\frac{e^2}{8 \pi^2} F_{\mu\nu}\tilde{F}^{\mu\nu}(y)\,,
\end{align}
where $\tilde{F}^{\mu\nu}\equiv \epsilon^{\mu\nu\alpha\beta} F_{\alpha\beta}/2$.  \Eq{eq:currentexpectation} can be reexpressed, using the canonical transformation to Grassmannian bilinears, as the second equation in \Eq{eq:Abelian-currents} while, as we have seen, the derivation of \Eq{eq:currentenonconservation} is quite nontrivial.

In some of the condensed matter literature, the emergence of the chiral anomaly is identified with the compressibility of phase space in the presence of a Berry monopole~\cite{Berry:1984jv}. However, as 
we have shown, in our construction, phase space is incompressible and the derivation of the anomaly equation is distinct from these considerations. Indeed, we showed previously that in a Euclidean formulation of the world-line path integral, Berry's phase is derived from the real part of the effective action; the anomaly, in contrast, arises from the imaginary phase that is made manifest by introducing the auxiliary gauge field $B_\mu$~\cite{Mueller:2017lzw}. Note that a similar observation concerning the connection between Berry's phase and anomalies was discussed recently in~\cite{Deguchi:2005pc,Fujikawa:2005tv,Fujikawa:2005tv,Fujikawa:2005cn,Fujikawa:2017ych}. In appendix \ref{app:Berry}, we will discuss further formulations of chiral kinetic theory including a Berry term and its interpretation in terms of the chiral anomaly.  
\section{Summary And Outlook}\label{sec:conclusions}
We presented in this manuscript a first principles construction of classical phase space with color and spin internal 
symmetries employing the Schwinger-Keldysh generalization of
the world-line approach to many-body quantum field theory. In this path integral formalism,
internal symmetries are expressed by elements of a Grassmann algebra. We obtained the 
classical phase space limit by taking the saddle-point limit of the quantum mechanical 
SK world line path integral in the truncated Wigner approximation, which is also equivalent to the classical
statistical approximation in field theory. For $SU(N_c)$ color, we derived the quantum Wigner distribution whose dynamics is defined by classical Poisson brackets and 
a Grassmann valued phase space measure. 

A canonical coordinate transformation connects our approach to
that of  Alekseev, Faddeev and Shatashvili \cite{Alekseev:1988vx} who used commuting Darboux variables to represent the color algebra.
The  underlying Grassmann algebra has a number of advantages. Without it, color phase space density 
and Liouville equation are decomposed into an infinite tower of equations, preventing any practical solution.
In contrast, we obtained a unique form for the color phase space distribution function, containing only singlet and octet components. Secondly, the action generating
the symplectic algebra for color using commuting Darboux variables in \cite{Alekseev:1988vx} contains
a topological Berry monopole which is problematic in the classical limit. In contrast, no such topological term arises in our approach and the classical phase space limit is 
conceptually clean. 

We discussed further the construction of classical phase space distributions and the corresponding phase space measure for spin-1/2 fermions. Starting from the SK world-line formulation,
we recovered the results of Berezin and Marinov \cite{Berezin:1976eg}. We explicitly demonstrated that one can write the most general form
for the Wigner phase space distribution function in terms of Grassmann spin bilinears and discussed the properties of the corresponding phase space measure. The phase space distribution is
parametrized by a pseudo-vector, which is the well-known Pauli-Lubanski vector satisfying the relativistic Bargmann-Michel-Telegdi equation. 

We proceeded to construct the vector and axial-vector currents of massless spin-1/2 fermions as phase space averages over the associated Liouville phase space distribution.
A naive representation of the latter shows that it is conserved. We showed that the nonconservation of this current due to the anomaly can be traced back to the proper gauge invariant regularization of the spectrum; this is shown to to be encoded in the initial conditions for the quantum Wigner distribution in the Schwinger-Keldysh representation. 

The results presented in this manuscript are part of a larger effort in constructing a consistent chiral kinetic theory.
In related work \cite{longpaper} in preparation, we derive a kinetic theory including collision terms from a fluctuation analysis of the Liouville equation using the relations derived here.
We have applied this formalism to the case of a nonequilibrium QCD plasma applying the consistent power counting pioneered by B\"odeker \cite{Bodeker:1998hm,Bodeker:1999ey}, Arnold, Son and Yaffe \cite{Arnold:2000dr}, and Moore \cite{Arnold:2002zm,Arnold:2003zc}. The physics goal of this specific application is to follow the spacetime development of the chiral magnetic current~\cite{Kharzeev:2007jp,Fukushima:2008xe} generated in the Glasma produced in ultrarelativistic heavy-ion collisions~\cite{Lappi:2006fp,Gelis:2006dv}, from the initial generation of net topological charge by sphaleron transitions~\cite{Mace:2016svc}, to the transport of this charge in background electromagnetic fields all the way, and ultimately to quantitatively ascertain the physical consequences of the same in experimental observables. While anomalous hydrodynamics is a robust approach to describe the late time dynamics of the chiral current in heavy-ion collisions, the results are very sensitive to the initial conditions. The initial spacetime development of the chiral magnetic current vial sphaleron transitions can be followed using classical statistical methods~\cite{Mueller:2016ven,Mace:2016shq}; however the validity of this approach fails before hydrodynamic equilibrium is attained. It is at this step that chiral kinetic theory is crucial for the subsequent description of chiral transport in the topological QCD background. Our work develops the framework for the implementation of chiral kinetic theory in this context. 

Phase space formulations for internal symmetries are also important for future experimental
investigations of the spin structure of the proton. For the first time, the proposed Electron-Ion Collider (EIC) \cite{Accardi:2012qut} will allow a unique three-dimensional
tomography of the proton, thus extending previous one-dimensional parton distribution functions to fully five-dimensional
quantum Wigner distributions \cite{Ji:2003ak,Belitsky:2003nz,Hatta:2016dxp}. Specifically, the EIC will allow to measure the decomposition of the proton spin's spin into intrinsic and orbital angular momentum
contributions of its parton constituents~\cite{Lorce:2011kd,Lorce:2011ni,Lorce:2012jy,Leader:2013jra}.  Recently, there has been considerable progress in studies of polarized parton distributions at small x~\cite{Kovchegov:2015pbl,Kovchegov:2016weo}. An outstanding question in this regard is the role of the chiral anomaly~\cite{Altarelli:1988nr,Carlitz:1988ab,Jaffe:1989jz}. The world-line framework is ideally suited for this discussion. We have recently adapted the world-line framework to study unpolarized and polarized parton distributions at small x~\cite{Tarasov-Venugopalan-inpreparation} and plan to address this issues in the near future~\cite{Mueller-Tarasov-Venugopalan}. 

This formalism may also be useful in extending the regime of validity of effective field theory descriptions  \cite{McLerran:1993ni,Gelis:2010nm,Blaizot:2016qgz} of gluon saturation because this framework naturally includes a Wess-Zumino term~\cite{Hatta:2005wp,Hatta:2006ci}.
\section*{Acknowledgements}
R.V. would like to dedicate this work to the memory of the late Misha Marinov and to his outstanding contributions on this topic. We thank Paolo Glorioso, Andrey Sadofyev and Andrey Tarasov for discussions. 
We thank Dima Kharzeev for asking key questions that influenced this work. We would like to 
thank ITP Heidelberg for their kind hospitality during the completion of this work. 
The authors are supported by the U.S. Department of Energy, Office of Science, Office of Nuclear Physics, under contract No. DE- SC0012704, and within the framework of the Beam Energy Scan Theory (BEST) Topical Collaboration. NM is funded by the Deutsche Forschungsgemeinschaft (DFG, German Research Foundation) - Project 404640738.

%
%
\appendix

\section{Derivation of Liouville's theorem}\label{app:phasepsace}
In this appendix, we will derive Liouville's theorem from quantum phase space in the semi-classical approximation. Firstly, we demonstrate how \Eq{eq:Liouville1} follows from \Eq{eq:defW}.
We write \Eq{eq:defW} in the more compact form
\begin{align}W_A^\chi&[\bar{x}(\tau),\bar{p}(\tau),\bar{\lambda}(\tau),\bar{\lambda}^\dagger(\tau)] \nonumber\\
= & \int d^4\bar{x}_id^4\bar{p}_i d^4\bar{\lambda}_id^4\bar{\lambda}_i^\dagger \,W_A^\chi[\bar{x}_i,\bar{p}_i,\bar{\lambda}_i,\bar{\lambda}^\dagger_i] \nonumber\\
&\times\delta[x(\tau)-x_{cl}(\tau;x_i)] \,\delta[p(\tau)-p_{cl}(\tau;p_i)] \nonumber\\&\times\delta[\lambda(\tau)-\lambda_{cl}(\tau;\lambda_i)]\,\delta[\lambda^\dagger(\tau)-\lambda^\dagger_{cl}(\tau;\lambda^\dagger_i)]\,,
\end{align}
where $x_{cl}(\tau;x_i)$ denotes the solution to the classical equations of motion at time $\tau$ for given initial conditions $x_i$. Secondly, taking the derivative $d/d\tau$ yields Liouville's equation, \Eq{eq:Liouville1},
\begin{align}\label{eq:Liouvilledetails}
\frac{d}{d\tau} W_A^\chi =& \int d^4\bar{x}_id^4\bar{p}_i d^4\bar{\lambda}_id^4\bar{\lambda}_i^\dagger \,W_A^\chi[\bar{x}_i,\bar{p}_i,\bar{\lambda}_i,\bar{\lambda}^\dagger_i] \nonumber\\
&\times \left( \dot{x}_\mu\frac{\partial }{\partial x_\mu} +\dot{p}_\mu\frac{\partial }{\partial p_\mu}  +\dot{\lambda}_a\frac{\partial }{\partial \lambda_a}+\dot{\lambda}^\dagger_a\frac{\partial }{\partial \lambda^\dagger_a}  \right)
\nonumber\\
&\times\delta[x(\tau)-x_{cl}(\tau;x_i)] \,\delta[p(\tau)-p_{cl}(\tau;p_i)] \nonumber\\&\times\delta[\lambda(\tau)-\lambda_{cl}(\tau;\lambda_i)]\,\delta[\lambda^\dagger(\tau)-\lambda^\dagger_{cl}(\tau;\lambda^\dagger_i)]\nonumber\\
=& \left( \dot{x}_\mu\frac{\partial }{\partial x_\mu} +\dot{p}_\mu\frac{\partial }{\partial p_\mu}  +\dot{\lambda}_a\frac{\partial }{\partial \lambda_a}+\dot{\lambda}^\dagger_a\frac{\partial }{\partial \lambda^\dagger_a}  \right)W_A^\chi \,,
\end{align}
where we used the identity $(d/d\tau)\delta[x(\tau)-x_{cl}(\tau;x_i)]  = \dot{x}_\mu\, (\partial /\partial x_\mu) \delta[x(\tau)-x_{cl}(\tau;x_i)]$ and similar identities for $p,\lambda,\lambda^\dagger$. Alternatively, \Eq{eq:Liouvilledetails}
can be written as 
\begin{align}\label{eq:Liouvillecompact}
\frac{ dW_A^\chi}{ d\tau} = \{  W_A^\chi ,H_W\}\,,
\end{align}
where $H_W$ is the Weyl symbol of the Hamiltonian and $\{ \cdot,\cdot\}$ denote Dirac brackets. Fixing a specific world line parameterization, $\tau = \tau(x^0)$, allows one to write the
right-hand of \Eq{eq:Liouvilledetails} in non-covariant form,
\begin{align}
\Big( \partial_t - \dot{\mathbf{x}} \frac{\partial}{\partial \mathbf{x}}  -\dot{\mathbf{p}} \frac{\partial}{\partial \mathbf{p}}   -\dot{\lambda}_a\frac{\partial }{\partial \lambda_a}+\dot{\lambda}^\dagger_a\frac{\partial }{\partial \lambda^\dagger_a} \Big) W_A^\chi=0\,,
\end{align} 
which demonstrates that indeed $d W_A^\chi/d\tau=0$. The absence of an explicit $\tau$-dependence can alternatively be understood as a gauge symmetry related to reparametrization 
invariance on the worldline.

\section{Color phase space: Some identities}\label{app:color}
In this appendix, we will fill in some details of classical phase space using Grassmanian variables for color. First, \Eq{eq:ident1} is demonstrated through direct integration
\begin{align}\label{eq:ident1proof}
\int dQ \,Q^a Q^b &=
\int \mathcal{D}\lambda\,\mathcal{D}\lambda^\dagger \,( \lambda_c^\dagger t^a_{cd} \lambda_d)( \lambda^\dagger_e t^b_{ef} \lambda_f )\nonumber\\&= -\epsilon_{ceg}\epsilon_{dfg}\, t^a_{cd} t^b_{ef}=\text{Tr}(t^at^b)=\frac{1}{2}\delta^{ab}\,,
\end{align}
and similarly for \Eq{eq:ident2}
 \begin{align}\label{eq:ident2proof}
\int dQ \,&Q^a Q^b Q^c=
\int \mathcal{D}\lambda\,\mathcal{D}\lambda^\dagger  \,( \lambda_d^\dagger t^a_{de} \lambda_e)( \lambda^\dagger_f t^b_{fg} \lambda_g ) (\lambda^\dagger_h t^c_{hi} \lambda_i )\nonumber\\& = -\epsilon_{dfh}\epsilon_{egi} (t^a_{de}t^b_{fg}t^c_{hi})  = \text{Tr}(t^b\{t^a,t^c\}) = \frac{A_R}{2} d^{abc}\,.
\end{align}
To compute the coefficients $c_i$ in \Eq{eq:generaldist}, we take moments according to \Eqs{eq:mom0}{eq:mom3}. We obtain 
\begin{align}
f^a(x,P)&=\int dQ \,Q^a \, f(x,P,Q) \nonumber\\&= \frac{c_1}{2}f^a(x,P) + \frac{A_R}{2}c_2 f^{bc}(x,P) d^{abc}\,,\\
f^{ab}(x,P)&=\int dQ \,Q^a Q^b \, f(x,P,Q) \nonumber\\&= \frac{c_0}{2}\delta^{ab}\,f(x,P)+\frac{A_R}{2}c_1 d^{abc}f^c(x,P)\,,\\
f^{abc}(x,P)&=\int dQ\, Q^a Q^b Q^c\, f(x,P,Q) \nonumber\\&= \frac{A_R}{2} c_0 d^{abc} \,f(x,P)\,,\\
f(x,P)&=\int dQ\, f(x,P,Q)=\frac{A_R}{2} c_3 d^{abc} f^{abc}\,,
\end{align}
so that  $ c_1=2$, $c_2=0$, $c_3 = 4/(A_R^2\,c_0 d^2)$ and $c_0=1$. In the context of Liouville's equation, one encounters derivatives with respect to color variables and we will find the following
identities to be useful:
\begin{align}
\int dQ Q^a \frac{\partial}{\partial Q^b} f(x,P,Q)&=\frac{3}{d^2}d^{acd}d^{bcd}\, f(x,P)\,,\\
\int dQ Q^a Q^b \frac{\partial}{\partial Q^c}   f(x,P,Q) &= \delta^{ab} f^c(x,P)\,.
\end{align}

\section{Compressibility of quantum phase space and the chiral anomaly}\label{app:Berry}
It is well known \cite{Moyal:1949sk} that the quantum mechanical formulation of phase space  violates Liouville's theorem by compressible corrections to \Eq{eq:Liouvillecompact} at $O(\hbar^2)$. These can be understood from 
Moyal's equation
\begin{align}\label{eq:Moyal}
\frac{ dW_A^\chi}{ d\tau} = -2 H_W \sin\left[ \frac{\Lambda}{2}\right] W_A^\chi = \{  W_A^\chi ,H_W\}\ + O(\hbar^2)\,,
\end{align}
with the Poisson/Dirac brackets as defined in \Eq{eq:Dirac} and the bilinear operator $\Lambda $ which satisfies 
\begin{align}
A\,\Lambda\, B\equiv \{ A,B\}\,.
\end{align}
In the more general formulation of \Eq{eq:Moyal}, one can systematically derive quantum corrections to the semi-classical approximations presented in this manuscript. 

One might also speculate about a connection between the quantum anomaly and (compressible) quantum connections in Moyal's equation.
We wish to argue that there is no such connection. In fact, one can demonstrate that the generalized coordinates $(x^\mu, p^\mu, \psi^\mu)$, introduceXkd in section \ref{sec:spin},
are canonical variables and that phase space is incompressible in the semi-classical limit.  Yet, in this limit the anomaly is manifest from the SK path integral, as discussed in section \ref{sec:anomaly}.

In this context, let us consider semi-classical phase space formulations including a Berry term \cite{Berry:1984jv,Littlejohn:1991zz,Sundaram:1999zz,Xiao:2005qw,Son:2012wh, Son:2012zy}, which involves
non-canonical variables and  compressible phase space in the semi-classical limit\footnote{A nice discussion may be found in \cite{Littlejohn:1991zz} of the role of Berry phases in
multicomponent WKB problems and its interpretation as a coordinate transformation to gauge invariant, but non-canonical phase space variables.}. Son and Yamamoto
proposed a semi-classical effective many-body theory for chiral fermions following the description of Xiao, Shi and 
Niu \cite{Xiao:2005qw}  for semi-classical Bloch electrons in weak electromagnetic fields. A similar discussion may also be found in Sundaram and Niu \cite{Sundaram:1999zz}. This theory is summarized in the effective classical equations,
\begin{align}\label{eq:defBerrytheory}
\dot{\mathbf{x}}& = \frac{1}{\hbar} \frac{\epsilon_n (\mathbf{p})}{\partial \mathbf{p}} - \dot{\mathbf{k}}\times \mathbf{\Omega}_n(\mathbf{p})\,,\\
\hbar{\dot{\mathbf{p}}} &= \-e\mathbf{E}(\mathbf{x}) - e\dot{\mathbf{r}}\times \mathbf{B}(\mathbf{x})\,,
\end{align}
where $\mathbf{\Omega}$  is the Berry curvature \cite{Berry:1984jv} and $\epsilon_n$ the energy of the $n$-th energy band. Xiao et al.~\cite{Xiao:2005qw} 
demonstrated that phase space is compressible in this theory, with the volume element
\begin{align}
\Delta V\equiv \frac{\Delta V_0}{1+e\mathbf{B}\cdot \mathbf{\Omega}}\,,
\end{align}
and $\Delta V_0\equiv d^3x d^3 p$ given in the absence of magnetic fields (in the asymptotic past). They further
suggested that the particle number density at zero temperature and finite chemical potential should be defined to be
\begin{align}\label{eq:electrondesnity}
n_e = \int^\mu \frac{d^3p}{(2\pi)^3}\,\left[ 1 + \frac{e\mathbf{B}\cdot \mathbf{\Omega}}{\hbar}\right]\,,
\end{align}
where the integrand includes modes with energy below the chemical potential $\mu$. 

Son and Yamamoto \cite{Son:2012wh, Son:2012zy} interpret \Eq{eq:electrondesnity}
as the chiral charge density and obtain their well known anomaly result by assuming constant $\mu$. 

In contrast, Xiao et al.~\cite{Xiao:2005qw} give a different interpretation of  \Eq{eq:electrondesnity}, where particle number is conserved. They note that the
 chemical potential is not constant in a magnetic field. Its time dependence causes a change in the fermi volume~\cite{Xiao:2005qw}, precisely compensating the correction term to the electron density in \Eq{eq:electrondesnity}. This interpretation is more in line with our understanding that the compressibility of phase space and the chiral anomaly have different origins. However it 
 is indeed remarkable that the expression obtained by Son and Yamamoto from the compressibility of phase space in the presence of a Berry term recovers precisely the 
 form of the anomaly. The puzzle this presents, and its definitive resolution, deserves further study. 

\bibliographystyle{apsrev4-1} 
\bibliography{references}

\end{document}